\documentclass[useAMS,usenatbib]{mn2e}
\usepackage{graphicx,times}
\usepackage{natbib}
\usepackage{amssymb}

\title[Simulations of gas sloshing in A2052]{Simulations of gas sloshing in galaxy cluster Abell~2052}

\author[R. E. G. Machado \& G. B. Lima Neto]{Rubens E. G. Machado\thanks{E-mail: rubens.machado@iag.usp.br} and Gast\~ao B. Lima Neto\\Instituto de Astronomia, Geof\'isica e Ci\^encias Atmosf\'ericas, Universidade de S\~ao Paulo, R. do Mat\~ao 1226, 05508-090 S\~ao Paulo, Brazil}

\begin{document}  

\date{Accepted 2014 December 13. Received 2014 November 18; in original form 2014 September 11}

\pagerange{\pageref{firstpage}--\pageref{lastpage}} \pubyear{2014}

\maketitle

\label{firstpage}

\begin{abstract}
The intracluster plasma of Abell 2052 exhibits in X-rays a spiral structure extending more than 250 kpc, which is comprised of cool gas. This feature is understood to be the result of gas sloshing caused by the off-axis collision with a smaller subcluster.
We aim to recover the dynamical history of Abell 2052 and to reproduce the broad morphology of the spiral feature.
To this end, we perform hydrodynamical $N$-body simulations of cluster collisions.
We obtain two regimes that adequately reproduce the desired features. The first scenario is a close encounter and a recent event (0.8 Gyr since pericentric passage), while the second scenario has a larger impact parameter and is older (almost 2.6 Gyr since pericentric passage). In the second case, the simulation predicts that the perturbing subcluster should be located approximately 2 Mpc from the centre of the major cluster. At that position, we are able to identify an observed optical counterpart at the same redshift: a galaxy group with $M_{500} = (1.16 \pm 0.43) \times 10^{13} M_{\odot}$.
\end{abstract}



\section{Introduction}
\label{sec:intro}

Galaxy clusters often exhibit signs of having been perturbed by encounters with subclusters, in agreement with our current understanding of hierarchical large scale structure formation.
The gas in the intracluster medium commonly bears the signatures of such disturbances in the form of shocks, cold fronts and other irregularities \citep[e.g.][]{Markevitch2007}.

Cold fronts are contact discontinuities, across which temperature and density are discontinuous, but pressure is not. A type of cold front with particularly interesting morphology are distributions of cool gas in the form of a spiral that stems from the cluster core and reaches out to large distances, sometimes hundreds of kiloparsecs \citep[e.g.][]{Rossetti2013,Walker2014}. Such spiral features have been observed in numerous clusters, Perseus \citep{Churazov2003} being one of the most prominent ones. \cite*{Lagana2010} studied a sample of nearby galaxy clusters and found half of them having signs of spiral structure. Abell~2029 also has an important spiral feature \citep*{Clarke2004}, which was recently found to stretch as far as out as 400~kpc at its fullest extent \citep{PaternoMahler2013}. Further examples include Abell~1644 \citep{Johnson2010}, Abell~496 \citep{Roediger2012b}, among others \citep*[see][]{Owers2009,Ghizzardi2010}. Such spiral structures have been studied even in groups \citep[e.g.][]{Randall2009, Lal2013, Gastaldello2013}.

These spiral cold fronts are understood to arise from gas sloshing triggered by an off-axis collision. The sloshing mechanism was proposed by \cite*{Markevitch2001} as a bulk motion of cool gas that is displaced from the cluster core. This might happen when the cluster's potential suffers a gravitational disturbance (a pull) from a passing subcluster. In this scenario, the resulting spiral is comprised of a dense, cool, low entropy gas that was removed from the cluster core. The interface between the cold spiral and the surrounding warmer gas is a contact discontinuity.

Hydrodynamical simulations have been employed to study the details of the sloshing mechanism (\citealp{AscasibarMarkevitch2006}; \citealp*{ZuHone2010}). Idealized simulations of binary cluster collisions are well suited to explore this process, by assessing the influence of initial parameters (such as mass ratio, relative velocity, impact parameter, gas content). \cite{AscasibarMarkevitch2006} showed that the off-axis passage of a small subcluster can set off gas sloshing and thus induce a long-lived spiral feature. Their simulations also indicated that non-head-on encounters are capable of giving rise to gas sloshing even if the perturber itself is gasless, underscoring the gravitational basis of this otherwise hydrodynamical phenomenon. Hydrodynamical -- and sometimes magnetohydrodynamical \citep*{ZuHone2011b} -- simulations have been used to explore the parameter space of such cluster mergers \citep[e.g.][]{ZuHone2011}, but also to model individual objects \citep{Roediger2011, Roediger2012b,Johnson2012}.

\subsection{Abell 2052}

Abell~2052 (hereafter A2052) is a nearby cool core cluster, at a redshift of $z= 0.03498 \pm 0.00022$ \citep{Smith2004}. The average temperature is $kT \sim 3$~keV, dropping to $\sim 1$~keV in the centre. The central cD galaxy hosts an active galactic nucleus (AGN), and the energy output of this radio source is responsible for evacuating the innermost gas, giving rise to bubbles and shocks on the scale of about 20~kpc \citep{Blanton2011}. 

A2052 had been observed in X-ray by \textit{Einstein} \citep*{White1997}, \textit{ROSAT} \citep{Peres1998}, \textit{Chandra} \citep*[][]{Blanton2003}, \textit{Suzaku} \citep{Tamura2008} and \textit{XMM-Newton} \citep{dePlaa2010}. More recently, with very deep \textit{Chandra} observations, \cite{Blanton2011} were able to identify a previously undetected spiral feature, which extends across more than 250~kpc, being considerably larger than the inner cavities. It corresponds to an excess in X-ray surface brightness, and it has lower temperature, lower entropy and higher metallicity than the gas in its immediate vicinity. Since these are the typical signatures of sloshed gas, the spiral feature in A2052 is interpreted as the outcome of an off-axis collision with a subcluster.

A2050 is in the footprint of SDSS DR9\footnote{Sloan Digital Sky Survey, \texttt{www.sdss.org}.}, with 104 galaxies with measured redshift in the region $6.4^\circ < \delta < 7.6^\circ$ and $228.6^\circ < \alpha < 229.8^\circ$ (J2000). The mean redshift of these galaxies is $z = 0.0347 \pm 0.0002$.

Our aim in this work is to propose a dynamical history of A2052, describing the detailed set of circumstances that caused the gas sloshing in this particular cluster. In order to achieve this, we perform a set o high-resolution hydrodynamical simulations in an attempt to recover the approximate spiral morphology of the cold front. This paper is organized as follows. In Section 2 we present the simulation techniques and initial conditions. In Section 3 we discuss the results of the two scenarios, including the issue of the optical counterpart. In Section 4 we summarize and conclude. Throughout this work we assume a standard $\Lambda$CDM cosmology with $\Omega_{\Lambda}=0.7$, $\Omega_{M}=0.3$ and $H_{0}=70 \,h_{70}$~km~s$^{-1}$~Mpc$^{-1}$.


\section{Simulation setup}

In order to model collisions giving rise to sloshed gas, we set up initial conditions representing isolated galaxy clusters. They comprise dark matter and gas and were created using the methods outlined in \cite{Machado2013}. The goal of these dedicated simulations is to attempt to reproduce features of A2052, particularly its spiral structure, as much as possible. This simulation approach relies on various simplifications, but the comparison between numerical results and observations allows for the study of plausible dynamical histories, giving us elements to better understand the process of structure formation in the Universe.


\subsection{Techniques and profiles}

We consider the collision of two spherically symmetric galaxy clusters in a similar manner to \cite{Machado2013}. The simulations were performed with the parallel SPH (smoothed particle hydrodynamics) code \textsc{gadget-2} \citep{Springel2005}, with a softening length of $\epsilon=5$~kpc. The ICM (intracluster medium) is represented by an adiabatic ($\gamma=5/3$) ideal gas, and cooling is disregarded in the simulations. Since galaxies account for a small fraction of the total cluster mass \cite[$\sim 3$ per cent, e.g.][]{Lagana2008}, their individual gravitational contribution may be safely overlooked (their mass is taken into account together with the collisionless particles). Due to the neglect of stars in the simulations, star formation and feedback are evidently not present. Magnetic fields are not expected to play a major role in determining the global cluster morphology, and are not taken into account either. The evolution of the system is followed for about 8~Gyr, but the relevant phases take place in a time-scale of only a few Gyr. Because of the small spatial extent of the system, cosmological expansion is ignored. Simulations were carried out on a 2304-core SGI Altix cluster.

The density profile adopted for the dark matter halo follows the \cite{Hernquist1990} profile:
\begin{equation}
\rho_{h}(r) = \frac{M_{h}}{2 \pi} ~ \frac{r_{h}}{r~(r+r_{h})^{3}} \, ,
\end{equation}
where $M_{h}$ is the total dark matter halo mass, and $r_{h}$ is a scale length. This resembles the NFW profile \citep*{NFW1997} except in the outermost parts.

For the gas distribution, a \cite{Dehnen1993} density profile is adopted:
\begin{equation}
\rho_{g}(r) = \frac{(3-\gamma)~M_{g}}{4\pi} ~ \frac{r_{g}}{r^{\gamma}(r+r_{g})^{4-\gamma}} \, ,
\end{equation}
where $M_{g}$ is the gas mass and $r_{g}$ is a scale length. With the choice $\gamma=0$, the resulting profile is similar to that of a $\beta$-model \citep{Cavaliere1976}. The temperature profile follows from the assumption of hydrostatic equilibrium.

Prior to the actual runs, each cluster is allowed to relax in isolation for a period of 5~Gyr. For further details on how the numerical realisations are created, see \cite{Machado2013} and references therein.
\begin{table}
\caption{Inital condition parameters.}
\label{tb:models}
\begin{center}
\begin{tabular}{l c c c}
\hline
          & cluster C1 & cluster C2 & cluster C3 \\ 
\hline
$M_{\mathrm{total}}$  & $2.6 \times 10^{14}~M_{\odot}$ &  $6.0 \times 10^{13}~M_{\odot}$&  $5.0 \times 10^{13}~M_{\odot}$ \\
$M_{200}$             & $1.6 \times 10^{14}~M_{\odot}$ & $4.1 \times 10^{13}~M_{\odot}$ & $3.6 \times 10^{13}~M_{\odot}$ \\
$M_{500}$             & $1.4 \times 10^{14}~M_{\odot}$ & $3.6 \times 10^{13}~M_{\odot}$ & $3.2 \times 10^{13}~M_{\odot}$ \\
$r_{200}$             & 1105~kpc                       & 700~kpc                        & 666~kpc                        \\
$r_{500}$             & ~768~kpc                       & 493~kpc                        & 472~kpc                        \\
$r_{h}$               & ~250~kpc                        & 120~kpc                        & 100~kpc                        \\
$r_{g}$               & ~550~kpc                        & 220~kpc                        & 200~kpc                        \\    
$f_{\mathrm{gas}}$    & 0.22                           & 0.16                           & 0.16                           \\
$f_{\mathrm{gas}}(<r_{500})$    & 0.13                           & 0.07                           & 0.07                  \\
\hline
\end{tabular}
\end{center}
\end{table}


\subsection{Initial conditions for A2052}

Our attempt to model A2052 is somewhat facilitated by the understanding that an off-axis encounter will not disrupt the cluster core as much as a head-on collision would. This means that we may use the current observed profiles as fair approximations to the initial conditions, assuming that the passage of a small subcluster should not substantially alter the azimuthally-averaged profiles of mass and temperature. This favourable circumstance holds for the major cluster, of course; for the subcluster initial conditions, no such immediate constraints are available, in principle.

By contrast, the task of modelling a head-on collision of a major merger would be made more difficult by the disruption of the cores. In that case, the final configuration would not offer a good estimate of the initial profiles and thus there would be too much freedom in the choice of initial parameters.

\begin{figure}
\centering \includegraphics[width=\columnwidth]{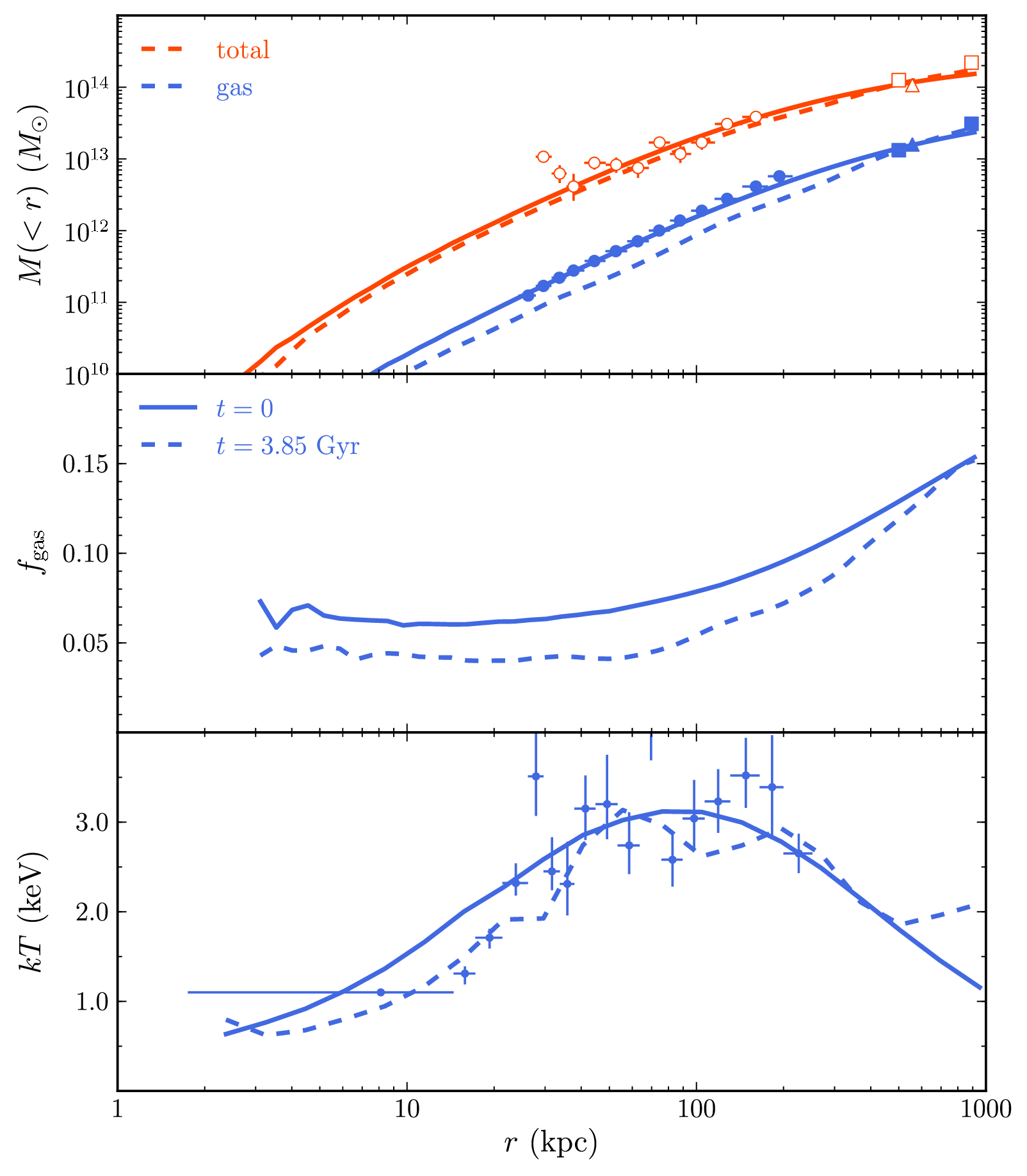}
\caption[]{Profiles of model A at $t=0$ Gyr (solid lines) and at $t=3.85$ Gyr (dashed lines). The upper panel displays the total (orange) and gas (blue) mass profiles, measured from the simulation. The data points (circles, squares and triangles) are taken respectively from \protect\cite{Blanton2003}, \protect\cite{White1997} and \protect\cite{Peres1998}. The middle panel gives the resulting gas fraction from the simulation. The lower panel compares the simulated temperature profile to the deprojected data points from \protect\cite{Blanton2003}.}
\label{fig:01}
\end{figure}

With this in mind, we use observational data to inform our estimate of initial parameters for the major cluster. We create cluster C1 (see Table~\ref{tb:models}) with parameters such as to approximately match the total mass profile, the gas profile and the temperature profile from \cite{Blanton2003} (with additional mass profile points from \citealt{White1997} and \citealt{Peres1998}). The initial conditions ($t=0$) of cluster C1 are the solid lines in Fig.~\ref{fig:01}. Table~\ref{tb:models} also gives the initial parameters of two subclusters that will be discussed in what follows.

We may try to evaluate the plausibility of these initial conditions with respect to scaling relations regarding mass, temperature and gas content. For example, given the mean temperatures $\langle T \rangle$ = 1.84, 0.88 and 0.82 keV (for clusters C1, C2 and C3 respectively), measured within $r_{500}$ in the initial conditions, we may estimate the masses using the $M_{500}-T_{X}$ scaling relation from \cite{Kravtsov2006}. These estimates result in approximately $M_{500} = 1.2 \times 10^{14}$, $3.8 \times 10^{13}$ and $3.4 \times 10^{13} ~M_{\odot}$ respectively, and they are roughly in the range of 5 to 15 percent of the masses used in the initial conditions (Table~\ref{tb:models}). With regard to the gas content, the results of \cite{Lagana2013} for example would suggest gas fractions of roughly 0.10--0.15 for cluster C1 and $\sim$0.07 for clusters C2 and C3, given their masses. These values refer to the region within $r_{500}$. In Table~\ref{tb:models} we provide the global gas fraction $f_{\rm gas}$, which takes into account merely the total masses of the Hernquist and Dehnen profiles. But we also provide the values of $f_{\rm gas}(<r_{500})$ measured from the initial conditions, which are in excellent agreement with what one would expect for objects of such masses. In any case, we note that -- at least for the cluster C1 -- the choice of parameters was constrained by the observed mass profiles and simultaneously by the observed temperature profile. In the cases of clusters C2 and C3, the initial conditions also very adequately follow the expected scaling relations, meaning that although we had no a priori information about these clusters, they were constructed with quite physically plausible parameters.

A large set of simulations was performed in order to explore the parameter space of possible collisions, but in this paper we will focus on describing and discussing the two scenarios that best match the X-ray observations, namely: \textit{model A} and \textit{model B}, each one a binary collision between a major cluster and a minor cluster. In both cases the total number of particles is $N=2.4\times10^6$, equally divided into dark matter and SPH particles.

In model A, the two clusters (C1 and C3) are set at an initial separation of 4000~kpc along the $x$-axis, and with an initial relative velocity of $-1000$ km/s in that direction. The impact parameter is $b=100$~kpc along the $y$ direction. In model B, the two clusters (C1 and C2) are set at an initial separation of 3000~kpc along the $x$-axis, and with an initial relative velocity of $-1000$ km/s in that direction. The impact parameter is $b=1000$~kpc along the $y$ direction. Note that the separation of the cluster centres at the instant of pericentric passage, $r_{\rm min}$, will be considerably smaller than the impact parameter $b$.

Thus, we have two regimes: model A is a close encounter with a mass ratio of about 4.5; while model B is a more distant encounter with a slightly more massive subcluster (mass ratio 3.9).

\begin{figure}
\includegraphics[width=\columnwidth]{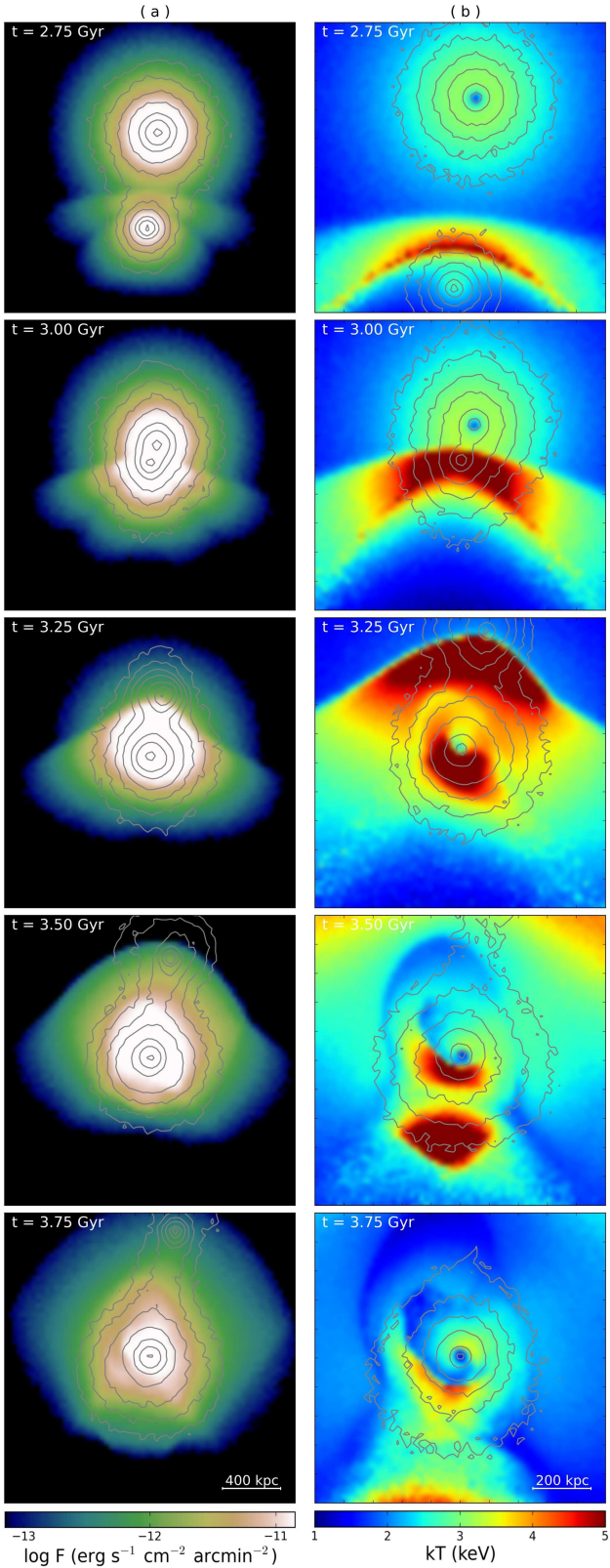}
\caption{For model A: (a) Time evolution of the X-ray surface brightness; (b) Time evolution of the emission-weighted projected temperature. In both columns, the contour lines represent the total (gas plus dark matter) projected mass. Note the different spatial scales: panels on the left column are 2000 kpc wide, whereas panels on the right column are 1000 kpc wide.}
\label{fig:02}
\end{figure}


\section{Results}

While the mass and temperature profiles of the major cluster are immediately -- and quantitatively -- constrained by observational data, the subcluster parameters are not, and neither is the initial configuration of the collision. We explored numerous variations of parameters -- such as mass ratio, impact parameter, velocities, and gas content of the subcluster -- within physically plausible ranges. We found two regimes of interest. In this paper we report on models A and B, representative of these regimes. In the following subsections we describe the features of these models, both of which have their strengths and shortcomings insofar as reproducing A2052 is concerned.

\subsection{Chandra observation}

We have downloaded ten publicly available \textit{Chandra} exposures of A2052: 9 by E.~Blanton (8 of which were observed in 2009 and one in 2006) and one pointing by C. Sarazin done in 2000. Following the ``Science Threads'' from the Chandra X-ray Center (CXC), using \textsc{ciao} 4.6\footnote{\texttt{asc.harvard.edu/ciao/}},
we have reprocessed these observations and merged them together producing a single broad-band (0.5--7.0 keV), exposure-map corrected surface brightness image with a pixel scale of $0.984$~arcsec (35/$h_{70}$ kpc at the cluster distance).

Using the \textsc{sherpa} package\footnote{\texttt{cxc.cfa.harvard.edu/sherpa/}}, we have fitted the flat x-ray image with a elliptically symmetric 2D $\beta$-model plus a flat background, masking the central region (dominated by the feedback of an AGN) and point sources. The best-fit $\beta$ model has $r_c = 48.13 \pm 0.13$~arcsec i.e., $(1684 \pm 5) h_{70}^{-1}$ kpc, $\beta = 0.44 \pm 0.01$, and an ellipticity of 0.18 with position angle equal to $0.8^\circ$ with respect to the North.

A residual image was produced by subtracting the symmetrical $\beta$-model from the original image. We will show this image bellow when exposing the simulations results.


\subsection{Model A: close encounter}

\begin{figure}
\centering \includegraphics[width=\columnwidth]{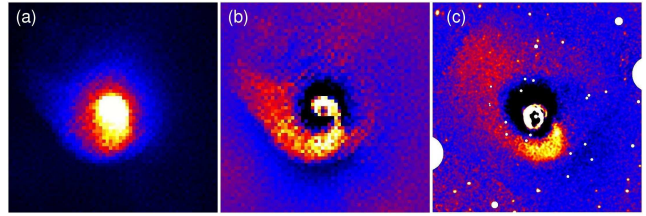}
\caption{For model A: (a) Simulated X-ray surface brightness map of model A; (b) Residuals from the subtraction of a fitted $\beta$-model from the simulated X-ray emission; (c) Residuals from the observational data (\textit{Chandra}). Each frame is 500 kpc wide.}
\label{fig:03}
\end{figure}

In model A, the simulation starts with clusters having an impact parameter of $b=100$~kpc. The instant of pericentric passage occurs at $t=3.05$~Gyr, when the separation is only $r_{\rm min}=42$~kpc. Despite this apparently small pericentric separation, the asymmetry if sufficient to trigger a substantial spiral of cold gas, which extends for more than 200~kpc.

In Fig.~\ref{fig:02} the time evolution of the X-ray emission is seen in snapshots at intervals of 0.25 Gyr (left-hand column). Likewise, the emission-weighted projected temperature is shown at the same intervals (right-hand column). Note, however, the different spatial scales meant to highlight the relevant features in each case. In both columns the overlaid contours represent the total projected mass, dominated by dark matter. The plane of the orbit is taken to be perpendicular to the line of sight. Projecting the simulations under different inclination angles did not improve the morphology, in comparison to the observed X-ray surface brightness of A2052.

\begin{figure}
\centering
\includegraphics[width=0.83\columnwidth]{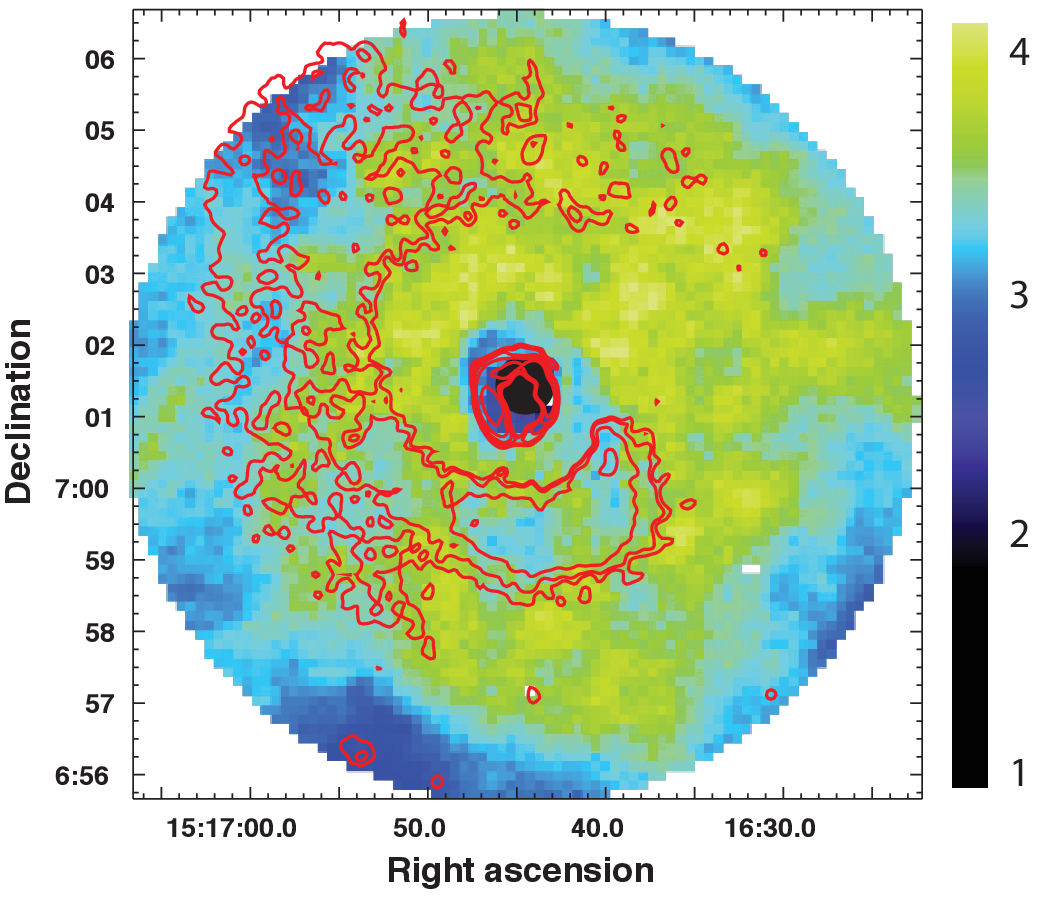}
\includegraphics[width=0.9\columnwidth]{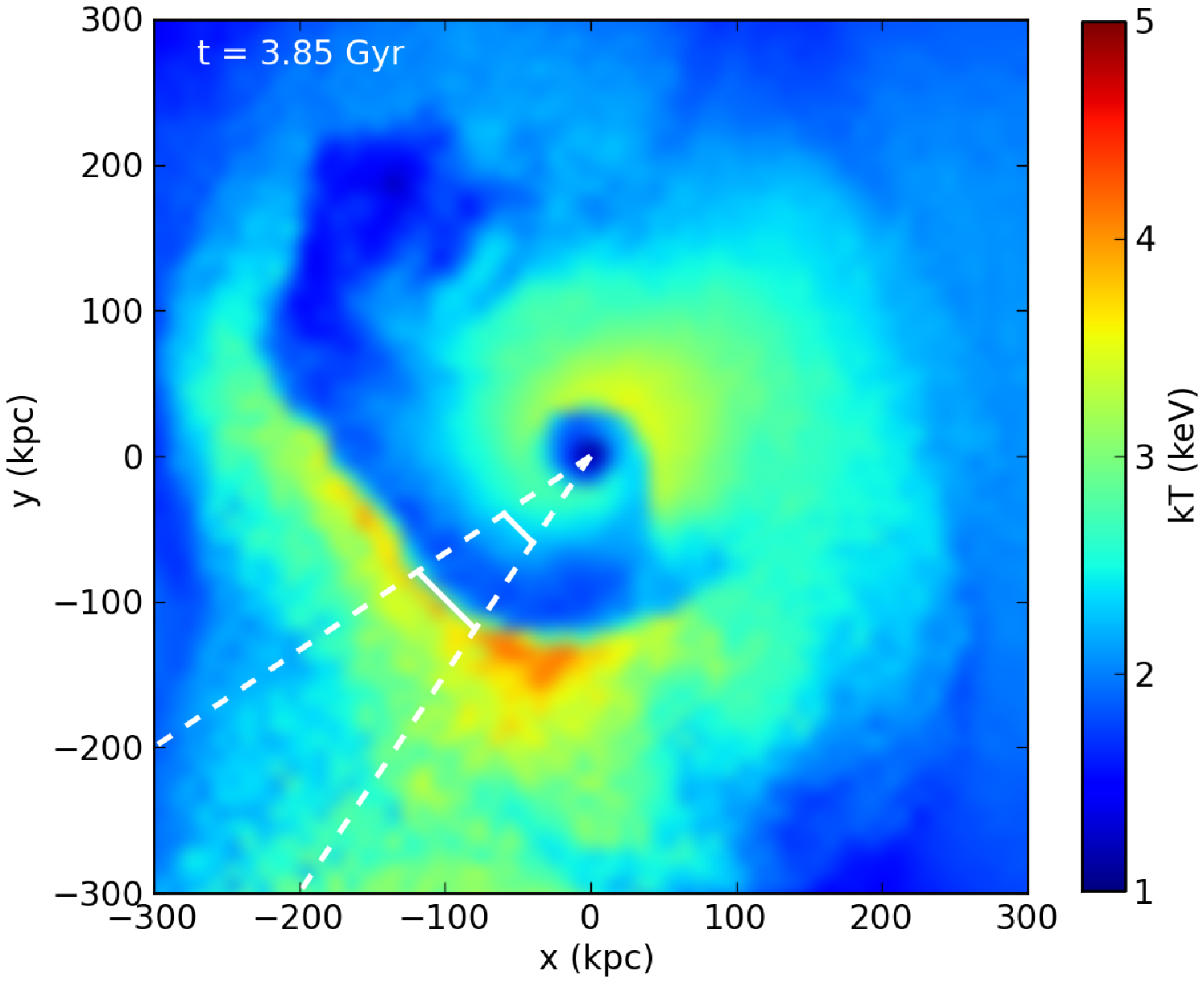}
\caption{Top: Observed temperature map of A2052. This figure was taken from the paper of {\protect \cite{Blanton2011}} and adapted. The contour lines represent the excess in X-ray emission. Bottom: Simulated emission-weighted projected temperature of model A at \mbox{$t=3.85$~Gyr}. The dashed sector represents in projection the cone within which the quantities of Fig.~\ref{fig:05} were measured.}
\label{fig:04}
\end{figure}

\begin{figure}
\centering
\includegraphics[width=0.9\columnwidth]{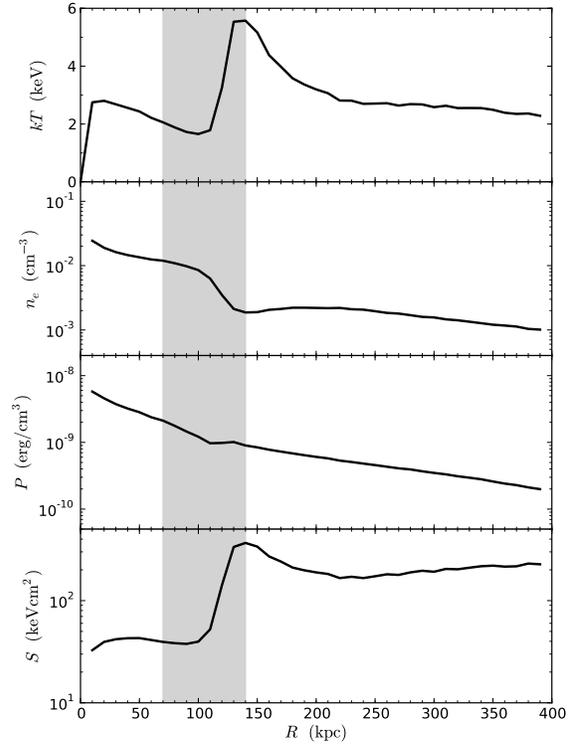}
\caption{For model A, the four panels display the temperature, electron number density, pressure and entropy profiles, measured within the region specified in Fig.~\ref{fig:04}. The shaded areas mark the width of the spiral.}
\label{fig:05}
\end{figure}

We searched for the instant when the morphology of the spiral feature best reproduces the observations. This instant of best match takes place approximately at $t=3.85$~Gyr, i.e.~0.8 Gyr after the pericentric passage. This snapshot is selected by inspecting neither the temperature map nor the X-ray emission, but the \emph{excess} in X-ray emission.

From the simulation output, first we produce a map of simulated X-ray emission \citep[for details, see][]{Machado2013} and then treat it as observational data: we fit a $\beta$-model, and subtract the fit from the emission map. The resulting X-ray excess is most prominently evident in this residual map. Figure~\ref{fig:03} compares simulated and observed X-ray excess (described in the previous section).

One notices that at this instant ($t=3.85$~Gyr), the size of the spiral feature is recovered, as well as its general shape and orientation. In particular, the inner portion of the spiral shows more intense emission.

\begin{figure}
\centering
\includegraphics[width=\columnwidth]{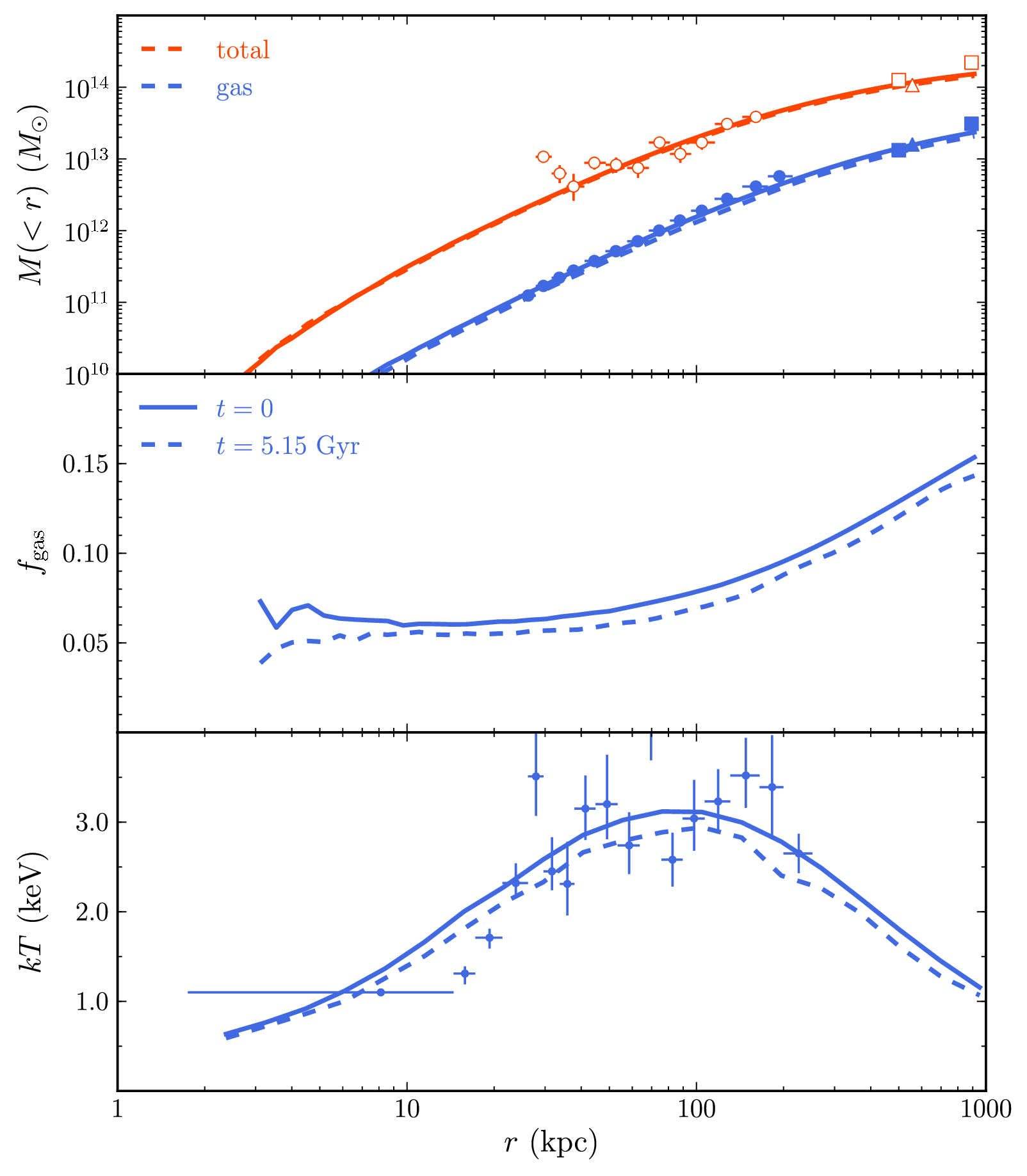}
\caption{Similar to Fig.~\ref{fig:01}, but for model B. Here, the dashed lines correspond to $t=5.15$~Gyr.}
\label{fig:06}
\end{figure}

The temperature profile of the cluster at $t=3.85$~Gyr is seen as the dashed line in the bottom panel of Fig.~\ref{fig:01}. The departure from the initial profile is not large; both profiles go from an average of 3 keV to about 1 keV in the centre. But being azimuthally-averaged, that radial profile conceals the details of the cold spiral. To examine in more detail the temperature structure of the gas, we display the best instant in the lower panel of Fig.~\ref{fig:04}, where a conic region is overlaid in projection onto the temperature map. This is the cone within which the thermodynamic quantities is Fig.~\ref{fig:05} were measured. The solid lines across the circular sector in Fig.~\ref{fig:04} -- marking the inner and outer faces of the spiral -- correspond to the edges of the shaded areas of Fig.~\ref{fig:05}. This analysis shows that the spiral is composed of gas which is cooler than its immediate vicinity, and it also has lower entropy. Here we adopt the usual proxy for entropy, $S = k~T~n_{e}^{-2/3}$, in terms of the electron number density $n_{e}$, temperature $T$ and Boltzmann constant $k$. At the outer face of the spiral, the noticeable drop in density is not accompanied by a relevant discontinuity in pressure.

The comparison between the simulated and observed temperature maps can be seen in Fig.~\ref{fig:04}, where the upper panel was taken from \cite{Blanton2011} and modified to a different color coding. The general relevant feature is that the gas within the spiral is somewhat cooler than its immediate vicinity. This property is strikingly more noticeable in the simulation. When considering such comparisons, one should bear in mind that our simulations do not include physical processes such as cooling or AGN feedback, and that they are idealised models in various other regards. The observed temperature map in this case does not afford a fruitful quantitative comparisons to the simulated temperature. We faced a similar situation in our simulations of A3376 \cite{Machado2013}, where the observed temperature map provided at most a qualitativey agreement, ensuring that the temperatures were at least roughly in the adequate range, but excluding more detailed comparisons. Here, a much more robust comparison is found with the maps of X-ray excess.

\begin{figure}
\centering
\includegraphics[width=\columnwidth]{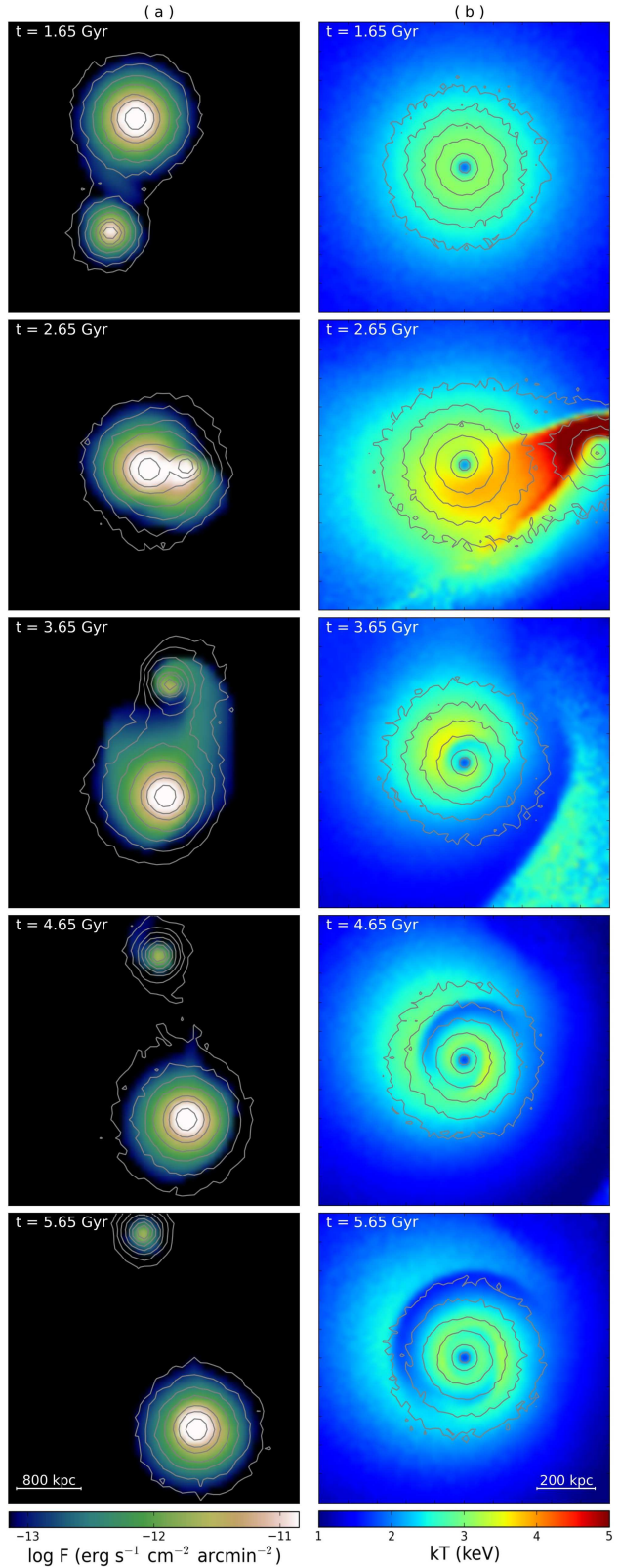}
\caption{Similar to Fig.~\ref{fig:02}, but for model B. Note the different spatial scales: panels on the left column are 4000 kpc wide, whereas panels on the right column are 1000 kpc wide.}
\label{fig:07}
\end{figure}

\begin{figure}
\centering
\includegraphics[width=0.65\columnwidth]{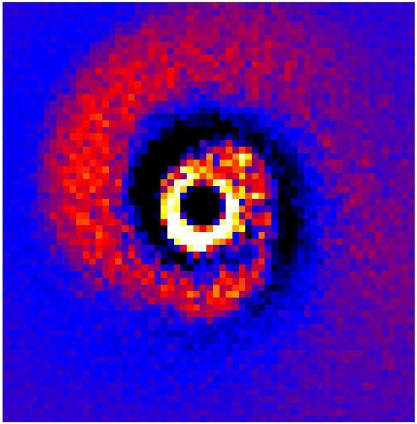}
\caption{Residuals of model B, after subtraction of a fitted $\beta$-model from the simulated X-ray emission. The frame is 500 kpc wide.}
\label{fig:08}
\end{figure}

Finally, it would be desirable to investigate the possibility of locating the perturbing subcluster. X-ray data is unavailable in the region and a search using SDSS DR9 data reveals no optical counterpart at the predicted location for model A. This issue is further discussed in Section \ref{sec:optical}.


\subsection{Model B: distant encounter}

In model B, the subcluster is slightly more massive, but the main difference with respect to model A is that here the impact parameter is an order of magnitude larger.

In model B, the simulation starts with clusters having an impact parameter of $b=1000$~kpc. The instant of pericentric passage occurs at $t=2.575$~Gyr, when the separation is $r_{\rm min}=446$~kpc. The instant of best match is found to be $t=5.15$~Gyr, i.e. about 2.575~Gyr after the pericentric passage. Note that the absolute timescales of models A and B are not directly comparable; not only because their dynamics are intrinsically different, but also because they do not start from identical separations. That being said, the physically meaningful comparison is between the times since pericentric passage.

The initial profiles of the major cluster are the same as in model A, but we see in Fig.~\ref{fig:06} that the final (azimuthally-averaged) profiles are even less perturbed here, because the subcluster passes at a greater distance. Nevertheless, a substantial spiral of cool gas develops as well in this case. In model B, it takes longer for the sloshing to set in, but the spiral persists for a longer time.

The global time evolution of the system is seen in Fig.~\ref{fig:07}, where X-ray emission and temperature are shown in snapshots 1.0 Gyr apart, and with spatial scales appropriate to exhibit the relevant features. As in model A, here too there is no inclination between the plane of the sky and the plane of the orbit.

At the instant of best match, the equivalent analysis is done to the simulated X-ray emission, whereby a fitted $\beta$-model is subtract from it. The residuals in Fig.~\ref{fig:08} show a spiral feature with the adequate size and orientation. Model B does not to reproduce the inner part of the spiral as well as model A does. However it succeeds in tracing the shape of the outer edges better than model A.

\begin{figure}
\centering
\includegraphics[width=0.9\columnwidth]{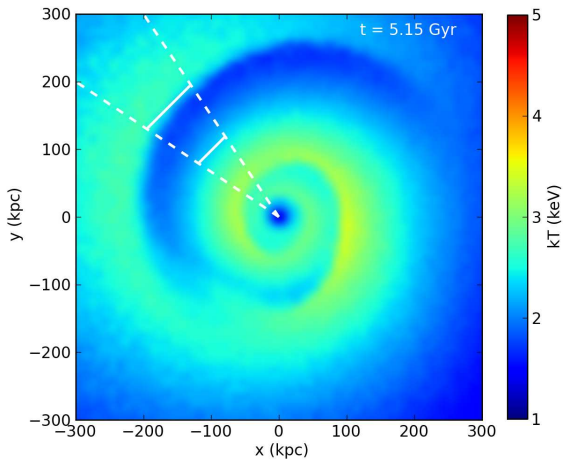}
\caption{Emission-weighted projected temperature of model B at \mbox{$t=5.15$~Gyr}. The dashed sector represents in projection the cone within which the quantities of Fig.~\ref{fig:10} were measured.}
\label{fig:09}
\end{figure}

\begin{figure}
\includegraphics[width=0.9\columnwidth]{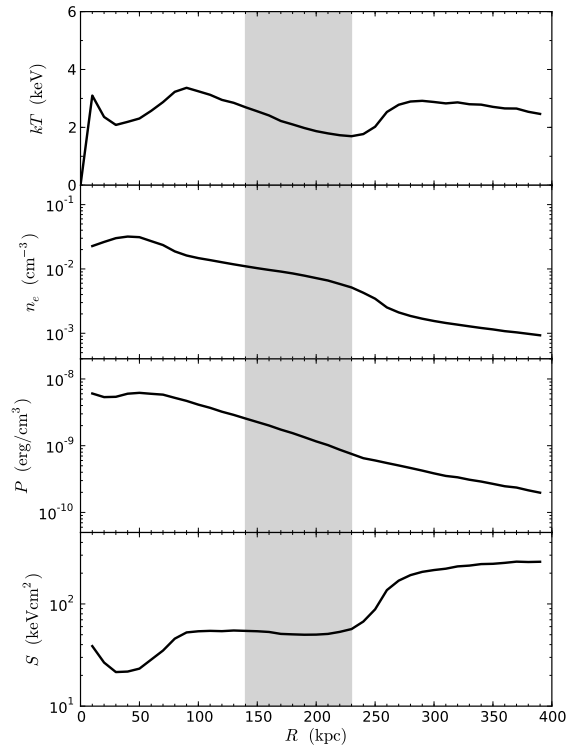}
\centering
\caption{For model B, the four panels display the temperature, gas density, pressure and entropy profiles, measured within the region specified in Fig.~\ref{fig:09}. The shaded areas mark the approximate width of the spiral.}
\label{fig:10}
\end{figure}

The azimuthally-averaged temperature profile is surely well matched (Fig.~\ref{fig:06}), but again we resort to measuring the profiles within a cone (Fig.~\ref{fig:09}) of the same aperture, pointing in a direction nor far from that used in model A. Qualitatively, the profiles of Fig.~\ref{fig:10} indicate again cool, low entropy gas within the spiral structure, although here the drops are considerably smoother, compared to model A profiles. The outer edge of the spiral -- where the temperature rises again -- corresponds to a very mild drop in gas density.

\subsection{Subcluster optical counterpart}
\label{sec:optical}

\begin{figure*}
\includegraphics[height=0.28\textwidth]{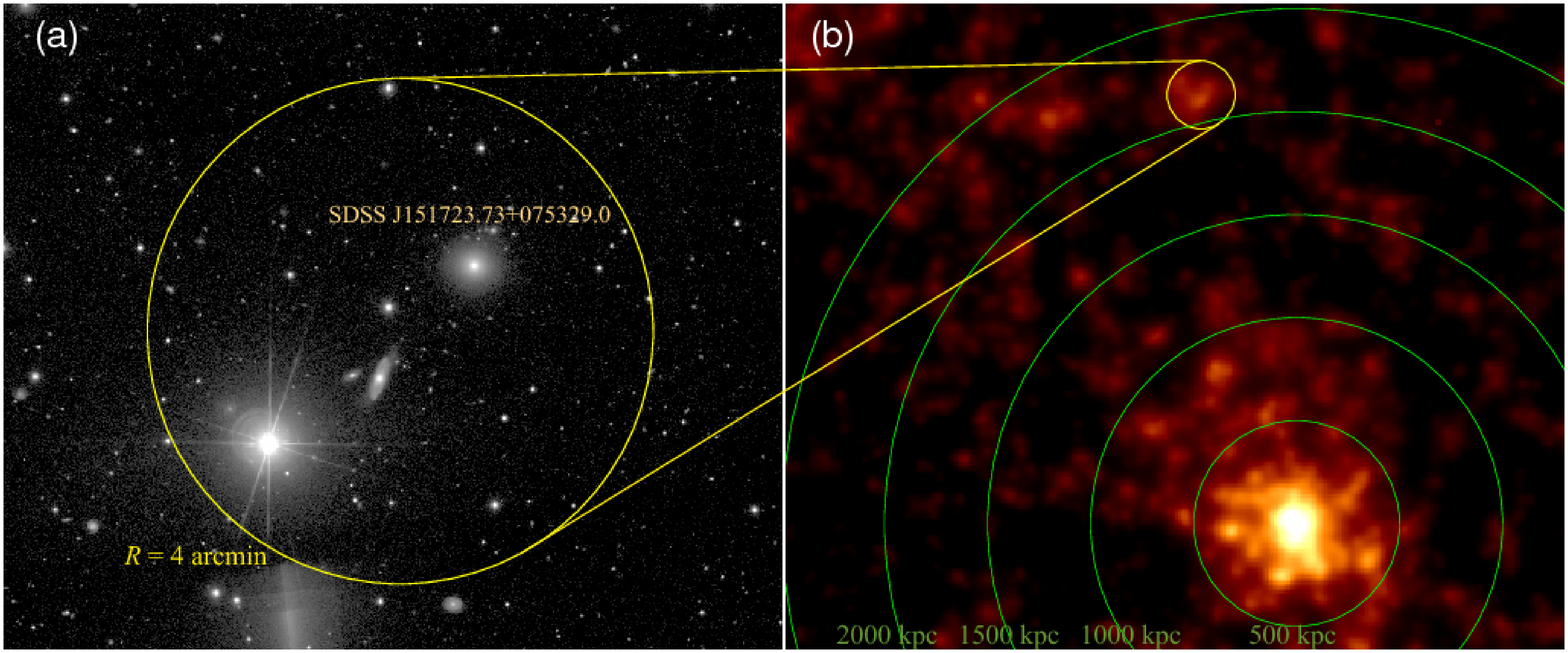}
\includegraphics[height=0.28\textwidth]{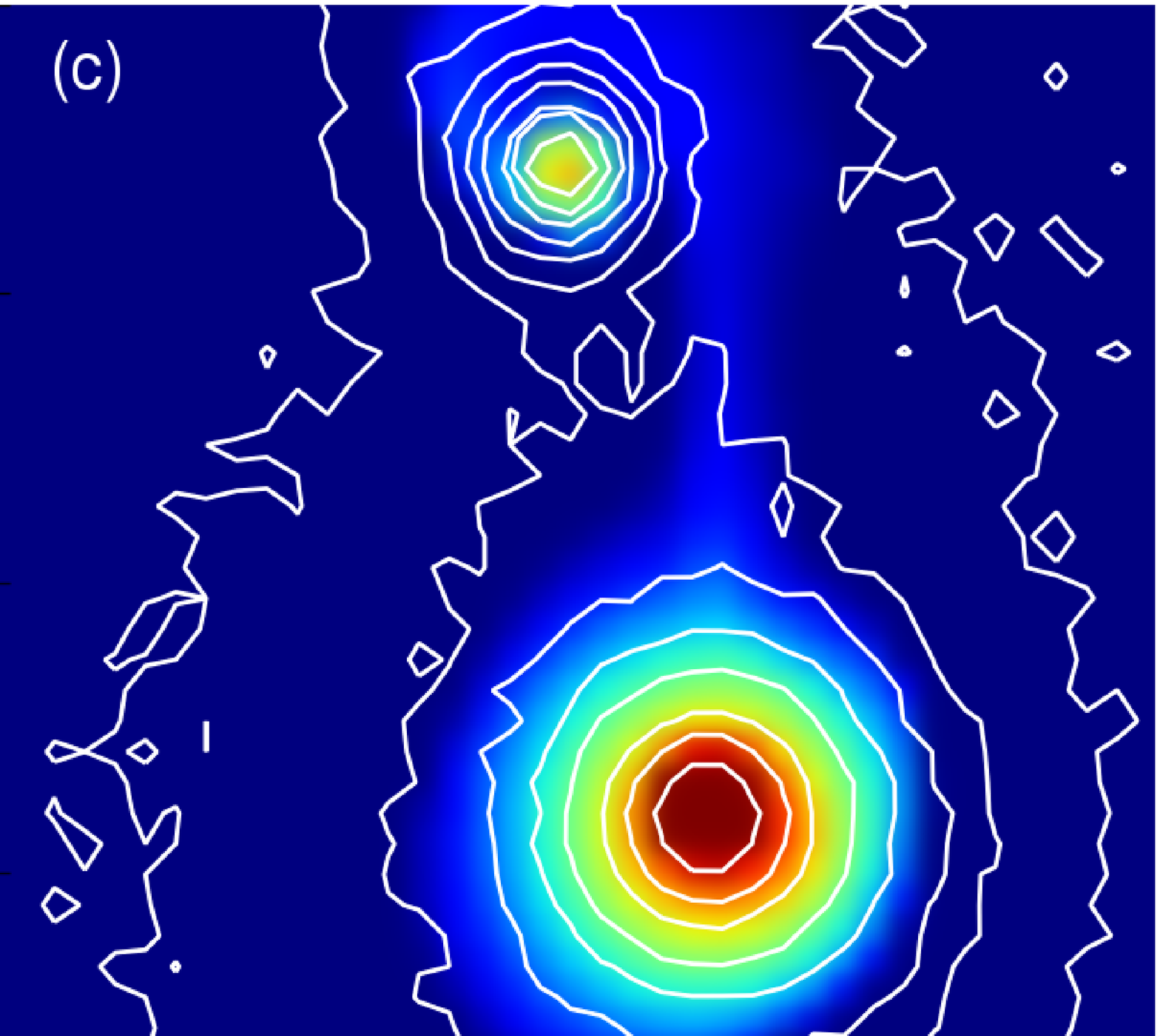}
\caption{(a) Zoomed-in view of the optical candidate; (b) Luminosity density in the $r$-band from SDSS galaxies; (c) Simulated X-ray emission (colours) with overlaid contours of total mass for model B.}
\label{fig:11}
\end{figure*}

A compelling prospect is the attempt to identify the perturbing subcluster in the observations. To evaluate this possibility we must consider the predicted location and properties of the subcluster in each case.

In model A, at the instant of best match, the subcluster is located 1020 kpc from the center of the major cluster, almost due north (in what would be the northwest quadrant). In the simulation, its position is easily identified due to the presence of its dark matter (the configuration is approximately that of the bottom panels of Fig.~\ref{fig:02}). However, after passing through the ICM of the major cluster, the subcluster will have lost part of its gas because of ram pressure. Furthermore, the virial radius of the subcluster is an ill-defined quantity at that time, because it is embedded within the mass of major cluster. Nevertheless we estimate roughly $5.4 \times 10^{12} M_{\odot }$ of gas contained in a region comparable to the subcluster's initial virial radius. This would give an indication of the gas mass one would have to expect, but it does not reflect the actual gas stripping, as most of the gas in that volume belongs to the major cluster itself. In fact, only $6.5 \times 10^{11} M_{\odot }$ of this gas belonged originally to the subcluster.

In model B, the subcluster is roughly twice as far, about 2120 kpc (in the northeast quadrant). By the time of best match, the subcluster has $M_{200} \sim 2.8 \times 10^{13} M_{\odot}$ and $r_{200} \sim 620$ kpc. This subcluster should also have suffered the effects of ram pressure, but to a lesser extent, as it did not pass through the densest inner parts. On the other hand, it has covered a longer path and is sufficiently outside the main cluster's virial radius. The result is that its gas mass has been reduced to about $1.6 \times 10^{12} M_{\odot }$, most of which ($1.2 \times 10^{12} M_{\odot }$) belonged originally to the subcluster itself. The temperature of this gas would be in the range of 0.6 to 0.8 keV, and the gas fraction is approximately 6 per cent within the subcluster's virial radius.

We should briefly mention that these mass determinations might be uncertain due the limitations of the SPH method employed here. There is a considerable debate concerning the accuracy of SPH in capturing certain phenomena related to fluid mixing instabilities \citep{Agertz2007}. Recent alternative SPH algorithms have been proposed to circumvent these difficulties \citep[e.g.][]{Hopkins2013}, but no such methods were used here.

Could either of these objects be identified from existing observational data? There are no \textit{Chandra} pointing reaching as far out as 1 or 2 Mpc away from the cluster centre. Even if there were, it is doubtful whether their X-ray emission would be easily detectable. The objects we seek would be in the mass range of galaxy groups, but with low gas content. In the case of model B, the simulated X-ray emission of the subcluster stands out noticeably against the background in the final configuration. The same cannot be said of model A. In the absence of X-ray data, we resort to SDSS optical data, from which we are able to obtain maps of galaxy luminosity density.

We have downloaded from the SDSS DR9 a table with all galaxies in a radius of 90~arcmin centred on the A2052 X-ray central peak, with magnitude $r > 12.6$~mag (corresponding to the brightest cluster galaxy) and $r < 23$~mag. For those galaxies with spectroscopic redshift, we retained only the ones with $0.001 < z_{\rm spec} < 0.1$, and for those objects without spectroscopy we used the photometric redshift, keeping only galaxies with $0 < z_{\rm phot} < 1.2$. With this selection, we filtered out most of the distant background galaxies while being very conservative and not excluding galaxies close (in redshift space) to the cluster.

We then produce a luminosity density map by smoothing and adding the luminosity of each galaxy on a coarse grid (1 pixel = 15 arcsec). The galaxy luminosity is smoothed following a projected 2D profile $\propto r^{-3}$ and scale length proportional to their effective radius (Fig.~\ref{fig:11}).

Inspection of the resulting map of luminosity density reveals no noticeable excess at the predicted subcluster location for model A. In fact, that region is even underdense.

On the other hand, at the predicted subcluster location for model B, a clear overdensity is seen (Fig.~\ref{fig:11}b). Moreover, we are able to pinpoint the galaxies responsible for this luminosity overdensity. They constitute a galaxy group at the same redshift as A2052. 

If these galaxies, having redshifts nearly identical to that of A2052, were located within the cluster's virial radius, they might more reasonably be understood to be mere cluster members. The fact that they lie more than 2 Mpc away from the centre of A2052 suggests that they should indeed be regarded as a separate group. The proximity of the redshifts also indicates that there is very little relative velocity between the cluster and the group along the line of sight. If this group can be interpreted as being the disturbing object that triggered the sloshing, then this holds an additional constraint: it reinforces the assumption that the orbit is seen on the plane of the sky, or at a very low inclination.

\subsection{Mass of the galaxy group}

\begin{table*}
\caption{Properties of the group galaxies: name, redshift, apparent magnitude, absolute magnitude, and luminosity (in the SDSS $i$-band).}
\label{tb:group}
\begin{center}
\begin{tabular}{l c c c c}
\hline
         Object name      &   $z$    &  $i$  &      $M_{i}$     & $L~(10^{10} L_{\odot})$    \\
\hline
CGCG 049-101              & 0.03574  & 13.12 &  -22.8 $\pm$ 0.1 &  8.30 $\pm$ 0.74 \\
CGCG 049-102              & 0.03623  & 14.08 &  -21.8 $\pm$ 0.1 &  3.43 $\pm$ 0.31 \\
2MFGC 12361               & 0.03436  & 15.36 &  -20.5 $\pm$ 0.1 &  1.05 $\pm$ 0.09 \\
SDSS J151719.66+073648.6  & 0.03375  & 15.75 &  -20.1 $\pm$ 0.1 &  0.74 $\pm$ 0.07 \\
2MASX J15182750+0750205   & 0.03373  & 15.90 &  -20.0 $\pm$ 0.1 &  0.64 $\pm$ 0.06 \\
2MASX J15173172+0804320   & 0.03737  & 15.96 &  -19.9 $\pm$ 0.1 &  0.61 $\pm$ 0.05 \\
SDSS J151729.92+081039.3  & 0.03330  & 16.31 &  -19.6 $\pm$ 0.1 &  0.44 $\pm$ 0.04 \\
SDSS J151757.42+080853.3  & 0.03483  & 16.55 &  -19.3 $\pm$ 0.1 &  0.35 $\pm$ 0.03 \\
SDSS J151804.11+080706.3  & 0.03216  & 16.80 &  -19.1 $\pm$ 0.1 &  0.28 $\pm$ 0.02 \\
SDSS J151757.48+074536.3  & 0.03248  & 17.03 &  -18.9 $\pm$ 0.1 &  0.23 $\pm$ 0.02 \\
\hline
\end{tabular}
\end{center}
\end{table*}

In the region of the overdensity seen in Fig.~\ref{fig:11} panel b, we were able to locate ten galaxies in the redshift range between 0.032 and 0.038. These are the objects that have SDSS spectra and are located within a circle of 16 arcmin radius. Their redshifts and magnitudes in the $i$-band are listed in Table~\ref{tb:group}.

The group mean redshift is $\overline{z} = 0.0344 \pm 0.0016$, which from the standard cosmology implies a distance of $d = 151.2 \pm 6.8$~Mpc and a distance modulus of $(i-M_{i}) = 35.9 \pm 0.1$~mag. We then compute the absolute magnitudes and corresponding luminosities in the $i$-band (Table~\ref{tb:group}).

To estimate the stellar mass of each galaxy, we used mass-to-light ratios as a function of absolute magnitude in the $i$-band, from \cite{Kauffmann2003}. The sum of the stellar masses amounts to $(2.69 \pm 0.59) \times 10^{11} M_{\odot}$. However, this value takes into account only the ten galaxies for which data is available. Even though these are expected to be the most luminous galaxies, a correction should be made for the fact that we are missing low-luminosity galaxies.

From the histogram of the galaxy luminosities, the magnitude completeness limit of the sample is estimated to be $M_{i} = -19.50 \pm 0.25$~mag, which is more conservative than the corresponding SDSS spectroscopic completeness at $r=17.8$~mag, \citep{Strauss2002}. We may compute the integrated luminosity, truncated at the corresponding $L_{min}$. The ratio of this quantity to the total integrated luminosity gives an estimate of the factor $f_{L}$ by which the total stellar mass is underestimated:
\begin{equation}
f_{L} = \frac{\int_{L_{min}}^{\infty} L~\phi(L)~\mathrm{d}L }{\int_{0}^{\infty} \quad L~\phi(L)~\mathrm{d}L} \, ,
\end{equation}
where $\phi(L)$ is the usual \cite{Schechter1976} luminosity function, with an appropriate constant $\alpha = -1.25$ \citep[e.g.][]{Propris2003}, and with the constant $L^{*}$ obtained from converting the $B$-band \mbox{$M^{*}=-19.5$}~mag into the $i$-band \citep{Fukugita1995}. We obtain a fraction $f_{L} = 0.66 \pm 0.06$. Consequently, the completeness corrected total stellar mass of the group is reckoned to be $(4.07 \pm 0.97) \times 10^{11} M_{\odot}$.

Finally, the total group mass must be estimated from its stellar mass. For this, we use a stellar fraction of $f_{\rm star}= 0.035 \pm 0.010$ from \cite{Lagana2013}, which is adequate for this group mass and refers to $M_{500}$. The total group mass is then estimated to be:
\[
M_{500}^{\rm obs} = (1.16 \pm 0.43) \times 10^{13} M_{\odot} \, .
\]
The relevant sources of error are the mass-to-light ratio, the completeness correction, and the stellar fraction of groups. This result is to be compared to the total mass of the subcluster measured from the simulation. From model B at $t=5.15$~Gyr (seen in Fig.~\ref{fig:11} panel c) we obtain a total mass of $M_{500}^{\rm sim} = 2.43 \times 10^{13} M_{\odot}$ within the radius $r_{500} = 435$~kpc.


\section{Summary and conclusions}

From deep \textit{Chandra} observations of the galaxy cluster A2052, \cite{Blanton2011} identified a previously unseen spiral feature, about 250 kpc in extent. This structure in the intracluster medium, seen as an excess in X-rays, is comprised of dense, cool, low entropy gas. Presumably, cool gas residing at the bottom of the potential well must have been extracted after the cluster was gravitationally perturbed by the passage of a subcluster. The resulting gas motion is what is known as the phenomenon of sloshing, and it is understood to be the cause of the spiral feature.

Assuming that the current state of A2052 is the aftermath of an off-axis collision with a subcluster, we employed hydrodynamical $N$-body simulations to study this collision. These simulations are idealised binary collisions that rely on several simplifications. The focus is chiefly on the global dynamics (i.e. the overall mass distribution) and on the large scale ($\sim$100 kpc) morphology of the sloshed gas. Particularly, we neglect altogether the effects of the AGN, and thus the inner structure ($\sim$10 kpc) of cavities is entirely beyond the scope of our simulations.

Our goal was to try to recover the broad morphological features that are observed in A2052, mainly the spiral structure. In fact, some global features are matched by construction, such as the mass profile and the azimuthally-averaged temperature profile of the major cluster. These profiles were available from observational data and they were employed as approximations to the initial conditions. Due to the presumed non-frontal nature of the collision, it is reasonable to assume that the overall mass distribution will not be greatly affected. This assumption is indeed borne out by the simulation results. On the other hand, the development of other features -- crucially the size and shape of the spiral structure -- cannot be ensured a priori. To match these features, we had to explore numerous combinations of initial parameters, many of which fail to develop the target morphology.

While the virial mass of the major cluster is essentially given, the properties of the subcluster (in hindsight more adequately referred to as ``the galaxy group'') are more difficult to settle, since no observational constraints were available beforehand, and thus there was too much liberty in the choice of initial conditions. Nevertheless, we were able to find combinations of subcluster mass and impact parameter that give rise to the desired outcome.

The two scenarios discussed in this paper are representative of different physical regimes. In the first case (model A), we have a close encounter, meaning a smaller impact parameter; in the second case (model B) the impact parameter is larger. In the first scenario, we would have a recent event (0.8 Gyr since pericentric passage) and a relatively short-lived spiral, i.e. a narrow time span during which the simulated morphology is comparable to A2052. In the second scenario the encounter would have been older (2.6 Gyr since pericentric passage) and the spiral persists for a longer time. The mass ratio is roughly 1:4 in both cases. In face of the simplifications involved, the agreement between model A and the observed X-ray excess is quite acceptable. The main drawback of model B is that, in the inner part of the spiral, the intensity is not as well matched as in model A. Nevertheless the outer reaches of the spiral are more fully traced in model B.

The identification of a plausible optical counterpart for the subcluster makes a compelling case in favour of model B. From SDSS optical data, we are able to identify an excess in the luminosity density map, at the predicted location. It corresponds to a galaxy group, at the same redshift as A2052. Furthermore, once the mass of the group is estimated, it is found to be in agreement with the mass from the simulation. We interpret that galaxy group as being the perturber responsible for triggering the gas sloshing. In such circumstances, the perturbing subcluster is often presumed to have been dispersed by tidal disruption, and believed to be impossible to identify. To our knowledge, this is one of few cases \citep[such as Abell 1644;][]{Johnson2010} in which a good candidate has been found.

These simulations accomplish an acceptable agreement with observation, at least as far as the broad morphological features of the ICM are concerned. This suggests that the effects of an AGN are not expected to play a determining role in the development of the sloshing spiral in A2052. 

We proposed two plausible scenarios of off-axis collisions that may give rise to the spiral structure seen in A2052. The regime of an old and distant encounter is favoured, because it correctly predicts the location of a galaxy group, which we identify in the optical. X-ray observations in the region 2 Mpc from the centre of A2052 could conceivably lead to the detection of the intra-group gas.


\section*{Acknowledgements}

This work has made use of the computing facilities of the Laboratory of Astroinformatics (IAG/USP, NAT/Unicsul), whose purchase was made possible by the Brazilian agency FAPESP (grant 2009/54006-4) and the INCT-A. Support from FAPESP (2010/12277-9) and from CNPq is acknowledged. REGM thanks Tatiana Lagan\'a for calling his attention to a cluster which certainly presents some features of interest. The authors thank her, Hugo Capelato and Felipe Andrade-Santos for useful discussions.

This research has made use of the NASA/IPAC Extragalactic Database (NED) which is operated by the Jet Propulsion Laboratory, California Institute of Technology, under contract with the National Aeronautics and Space Administration.

Funding for SDSS-III has been provided by the Alfred P. Sloan Foundation, the Participating Institutions, the National Science Foundation, and the U.S. Department of Energy Office of Science. The SDSS-III web site is \texttt{http://www.sdss3.org/}. SDSS-III is managed by the Astrophysical Research Consortium for the Participating Institutions of the SDSS-III Collaboration including the University of Arizona, the Brazilian Participation Group, Brookhaven National Laboratory, Carnegie Mellon University, University of Florida, the French Participation Group, the German Participation Group, Harvard University, the Instituto de Astrofisica de Canarias, the Michigan State/Notre Dame/JINA Participation Group, Johns Hopkins University, Lawrence Berkeley National Laboratory, Max Planck Institute for Astrophysics, Max Planck Institute for Extraterrestrial Physics, New Mexico State University, New York University, Ohio State University, Pennsylvania State University, University of Portsmouth, Princeton University, the Spanish Participation Group, University of Tokyo, University of Utah, Vanderbilt University, University of Virginia, University of Washington, and Yale University.


\bibliographystyle{mn2e.bst}
\bibliography{a2052.bib}
\bsp

\label{lastpage}

\end{document}